\documentclass[twocolumn,showpacs,superscriptaddress,aps,prb]{revtex4}
\usepackage[english]{babel}
\usepackage{xspace}
\usepackage{times,amsmath,amsfonts,amsthm}
\usepackage{graphicx}
\usepackage{epsfig}
\usepackage{wrapfig}
\usepackage{dcolumn}
\usepackage{bm}
\usepackage{color}

\begin{document}

\title{Stress release mechanisms for Cu on Pd(111) in the submonolayer and monolayer regimes}

\author{
J. Jalkanen}
\affiliation{Department of Applied Physics and COMP
Center of Excellence, Helsinki University of Technology, FI-02015
TKK, Espoo, Finland } 

\author{
G. Rossi}
\affiliation{Department of Applied Physics and COMP
Center of Excellence, Helsinki University of Technology, FI-02015
TKK, Espoo, Finland } 

\author{
O. Trushin}
\affiliation{
Institute of Physics and Technology, Yaroslavl Branch,
Academy of Sciences of Russia, Yaroslavl 150007, Russia
}

\author{
E. Granato}
\affiliation{
Laborat\'orio Associado de Sensores e Materiais, Instituto National de Pesquisas Espaciais,
12201-970 S\~ao Jos\'e dos Campos, SP Brasil
}
\affiliation{
Department of Physics, P.O. Box 1843,
Brown University, Providence, RI 02912-1843
}

\author{
T. Ala-Nissila} 
\affiliation{Department of Applied Physics and COMP
Center of Excellence, Helsinki University of Technology, FI-02015
TKK, Espoo, Finland } 
\affiliation{
Department of Physics, P.O. Box 1843,
Brown University, Providence, RI 02912-1843
}

\author{
S.-C. Ying}
\affiliation{
Department of Physics, P.O. Box 1843,
Brown University, Providence, RI 02912-1843
}

\begin{abstract}

We study the strain relaxation mechanisms of Cu on Pd(111) up to the
monolayer regime using two different computational methodologies,
basin-hopping global optimization and energy minimization with a
repulsive bias potential. Our numerical results are consistent with
experimentally observed layer-by-layer growth mode. However, we find
that the structure of the Cu layer is not fully pseudomorphic even
at low coverages. Instead, the Cu adsorbates forms fcc and hcp
stacking domains, separated by partial misfit dislocations. We also
estimate the minimum energy path and energy barriers for transitions
from the ideal epitaxial state to the fcc-hcp domain pattern.
\end{abstract}

\date{November 10, 2009}


\maketitle

{\it Introduction. \--} Metallic surfaces and nanostructures are
essential systems for heterogeneous catalysis. Combination of two
metals can lead to significant improvements in the variability and
frequency of the reactions catalyzed. To prepare controlled
nanostructures, it is crucial to understand the growth and stability
of heteroepitaxial metal systems, in particular for close-packed
surfaces. On fcc(111) it has been found that to release the
stress, the overlayer can adopt several alternative strategies which
lead to a structure of domains separated by partial misfit
dislocations \cite{Carter95, Hamilton95,Gabaly05, Hwang97,
Figuera01, Schmid97, Gunther95}. 
The domains correspond to the two favorable sites, fcc and hcp.
This behavior is expected to be ubiquitous and should not 
dependent strongly 
on the interaction potentials or the 
overlayer-substrate mismatch
\cite{Ubiq,Hamilton95,Pushpa03}. 

An interesting system, 
which has a large mismatch of $-7.1\%,$ is Cu on Pd(111). 
Experimentally it has been found that the growth of Cu
is layer-by-layer at room temperature \cite{Liu99,Oral90,deSiervo05A}. 
In the submonolayer regime the substrate becomes gradually covered 
by 2D islands \cite{Liu99}. Although near 1 ML coverage there is 
some indication 
of dislocations \cite{Oral90,Liu99,deSiervo05B}, only at $2$-$3$ ML
the average lattice constant reduces to the relaxed value. This has
been interpreted as a transition to relaxed morphology
\cite{Oral90,deSiervo05A} leading to the conclusion that the
pseudomorphic structure is stable up to $2$-$3$ ML.

In this work we unravel the atomic level stress release mechanisms 
of Cu on the Pd(111) surface from the submonolayer up to one
monolayer regimes under conditions when no 
alloying takes place \cite{alloying}.
We employ two efficient computational methodologies,
namely basin-hopping global optimization and energy minimization
with repulsive bias potential. Our numerical results are consistent
with the experimentally observed layer-by-layer growth. However, we
find that the structure of the Cu adsorbate is not fully
pseudomorphic even at low coverages. Instead, the Cu overlayer forms
fcc and hcp stacking domains, separated by partial misfit
dislocations. We also estimate the minimum energy path and energy
barriers for transitions from the ideal epitaxial state to the
fcc-hcp domain pattern.

\begin{figure}[ht]
\epsfig{file=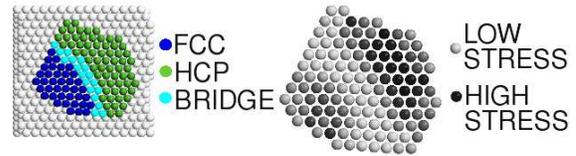, width=0.45\textwidth}
\caption{\label{fig:sm-submonolayer}
(Color online)
Left panel: Low energy structure from GO
with $N_{\rm Cu} = 120$. Pd atoms are white, while Cu 
colors
differ according to Cu stacking. Right panel: The map of atomic
stress in the same island. Stress is quantified as in Refs.
\onlinecite{Nielsen83,Vitek87,Ouahab05}. The fcc and hcp areas are under
tensile strain while the domain wall is nearly stress-free.}
\end{figure}

{\it Model and Methods. \--} In our calculations the system is
described as a stack of rectangular (111) layers whose shorter edges
are along close-packed rows in the $x$ direction, longer edges are
in the $y$ direction and $z$ axis is perpendicular to the (111)
plane. Periodic boundary conditions are applied in the $x$ and $y$
directions and two bottom substrate layers are fixed to the bulk
values. The number of atomic rows in the $y$ direction is chosen to
be twice that in the $x$ direction.
With this geometry the smallest simulation cell which can
accommodate layers with ideal lattice spacing of both Cu and Pd 
has $15\times30$ edge Pd atoms.

Metal interactions are modeled by the Embedded Atom Method (EAM)
\cite{Daw84,Foiles86}, which has given a good agreement with experiments
on the Cu/Pd(100) system \cite{Lu05}.
The energy of a Cu impurity in a Pd host is close to the
experimental value \cite{Foiles86} and the surface 
energies are consistent with the measurements 
\cite{Foiles86} and first-principles studies \cite{Chen03}.
The potential correctly predicts the system to be in the wetting regime,
although for a bulk alloy it favours fcc instead of
the observed B2 bcc phase \cite{Mottet02}.
To identify the low energy configurations, we employ two
complementary strategies, namely basin-hopping 
global optimization (GO) and activation-minimization (AM) techniques.
Both methods have been described in detail elsewhere
\cite{Wales03book,Rossi09,Jalkanen05}; thus we will only recite
their main features here.

{\it Global Optimization. \--} Basin-hopping global optimization
is a tool for finding, with fixed size and chemical
composition, the atomic 
configuration with the
lowest potential energy
\cite{Wales03book}.
The GO tool can start the search
from any given configuration, and at each step it proceeds as
follows: (i) The Cu atoms are perturbed by the \textit{shake} move
described in Ref. \onlinecite{Rossi09}, and alloying is prevented.
(ii) The perturbed system, including the Pd slab, is locally
minimized. (iii) The energy of the new local minimum is compared to
the previous one and the move is accepted according to the standard
Metropolis criterion.

Global optimizations for Cu island sizes $N_{\rm Cu} > 20$ start
from an initial, randomly-arranged Cu island locally minimized on
top of a $15\times30$ Pd slab with thickness varying between 3 and 6.
The two bottom layers are kept frozen in their ideal bulk
configuration, while the 
other layers are allowed to relax during the
local minimization procedure. One run of $10^{5}$ steps 
is performed for each of the systems considered. 

{\it Activation-Minimization Procedure. \--} The AM technique starts
from a perfectly pseudomorphic Cu overlayer. 
During both activation and energy minimization stages, the system follows 
the time-dis\-cre\-tized Newtonian equations of motion which are
solved with the standard leap-frog algorithm.
We perform the energy minimizations
with the Molecular Dynamics Cooling (MDC) technique\cite{Trushin09}.
On the activation stage we also apply MDC but now 
the system is under the influence of 
the repulsive bias potential (RBP)\cite{tru02b,tru04}.
In some cases, before turning on the RBP,
we increase the chance of getting certain kind of
defects by perturbing some selected atoms slightly
to a direction which is likely to initiate a nucleation process. 
After switching off the RBP we further relax the final state with MDC.
During activation, we save
intermediate states and use them as an input for the Nudged Elastic
Band (NEB) method to find the minimal energy path between the
initial and final states \cite{neb}.
Our final results were checked by varying the simulation
box size, Pd slab thickness, perturbations and the level of biasing and
our conclusions do not depend on these details either for AM or GO.

{\it Submonolayer regime. \--} In the submonolayer regime, we
performed GO for several island sizes ($5 < N_{\rm Cu} < 120$).
Consistently with the experimentally observed layer-by-layer growth
\cite{Liu99,deSiervo05A}, all the global minimum energy structures
for $N_{\rm Cu} > 5$ are flat. 
A low energy structure for $N = 120$ 
is shown in Fig.\ref{fig:sm-submonolayer}. It is worth noticing that the
adsorption energy of the island does not substantially depend on its
stacking. When the Pd slab has stacking sequence ABC, the Cu atoms
in the next layer can occupy sites A or B, which are referred to as
fcc or hcp stackings, respectively \cite{Carter95,Hamilton95}.  The
potential correctly assigns a lower cohesion energy to fcc than to
hcp bulk, although the difference is small, $E_{fcc-hcp} < 2$
meV/atom. The best adsorption site for a single Cu atom on Pd(111)
is on the fcc site, the energy difference from the hcp stacking
being $4.4$ meV. Consequently, islands with $N_{\rm Cu} < 41$ are
completely on the fcc sites. All the larger islands, where $60 \leq
N_{\rm Cu} \leq 120,$ are in both fcc and hcp stackings.

\begin{figure}[ht]
\begin{center}
\epsfig{file=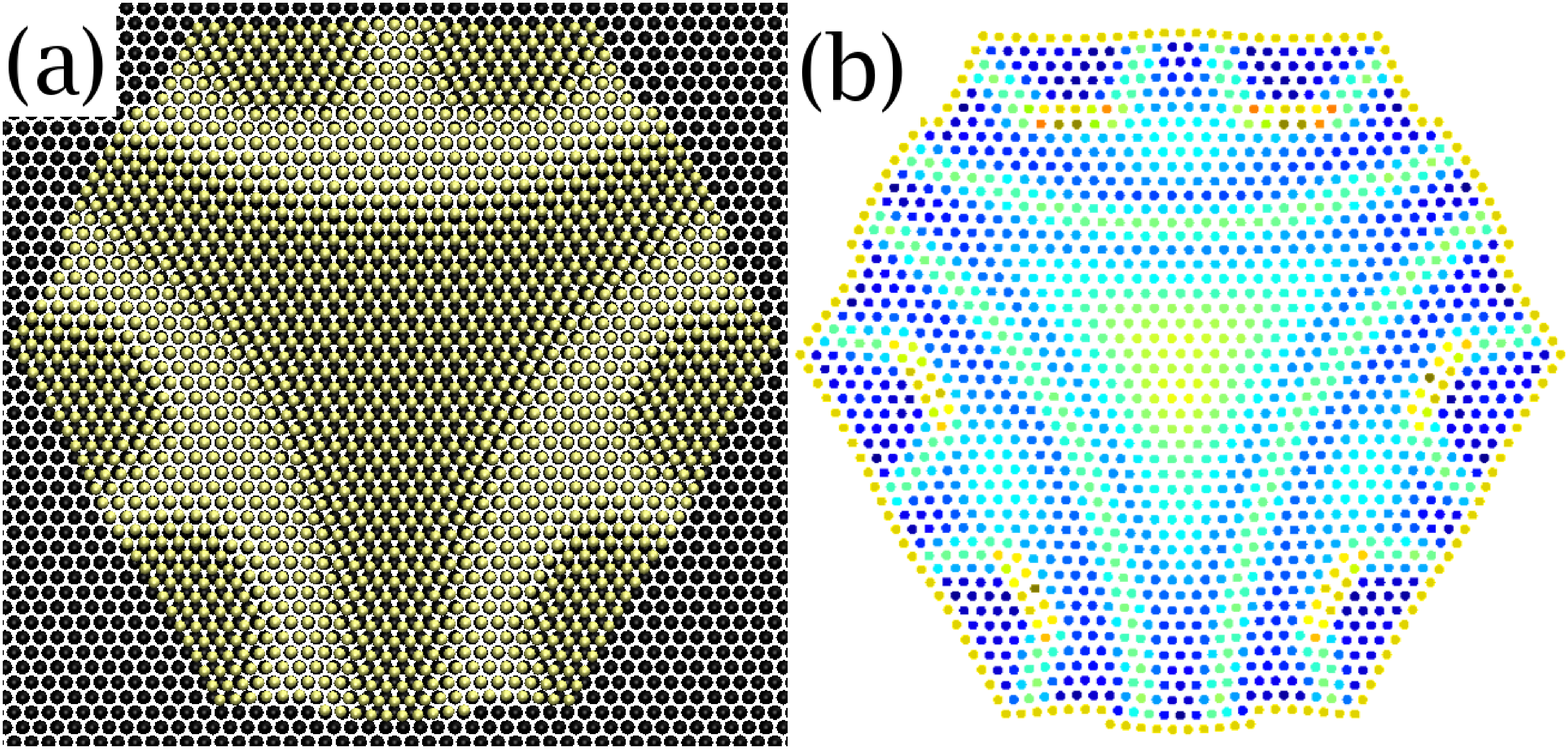, width=0.45\textwidth}
\caption{\label{fig:plates}
(Color online)
(a) shows a stacking domain trigon in a hexagonal Cu island
(light color) and the next-nearest substrate layer (dark color).
When the adsorbate occupies hcp sites, the dark substrate is 
occluded, otherwise the stacking is fcc-like and the substrate is 
visible through the overlayer. 
(b) shows the energy map for the trigon in (a).
The lightly colored high energy regions are in the same locations as the 
stacking domain walls of (a) and as the highly-stressed 
region at the center of the trigon. }
\end{center}
\end{figure}

A stress color-scale graph is shown in Fig. \ref{fig:sm-submonolayer}.
The stress of atom $i$ is quantified as 
the isostatic local atomic pressure\cite{Nielsen83,Vitek87,Ouahab05},
$ p(i)=-\sum_{j}r_{ij}\left(\partial E/\partial r_{ij}\right)/3 $
where $j$ runs over the neighbors of
atom $i,$ $E$ is the total potential energy of the system
and $r_{ij}$ is the distance between $i$ and $j$.
The tensile stress is largest near the center of the fcc
and hcp domains and stress release originates from
partial dislocations running along the 
domain walls.

In similar systems exhibiting 2D incommensurate ordering,
the walls are called heavy or superheavy, 
corresponding to atoms at neighbor fcc and hcp sites or atoms
on bridge positions, respectively \cite{Zeppenfeld88}. 
The latter are more common in islands.
In our system both walls are often distorted to bring atoms
closer to the equilibrium distance. Especially the 
superheavy walls are typically curved and extended over several 
lattice spacings.
While the wall atoms are at disadvantageous positions, the 
elastic relaxation, which would not be possible without the domain 
structure, has longer range
and releases energy in the whole island. Larger hexagonal islands
also display the domains in a pattern which is called a trigon
\cite{Hamilton95}. The center of the trigon shows energy increase
due to high stress levels, as shown in Fig.\ref{fig:plates}(b).

\begin{figure}[ht]
\epsfig{file=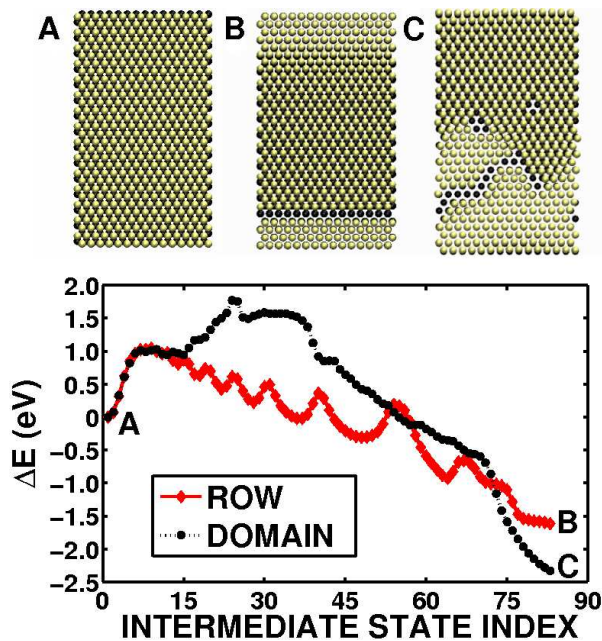, width=0.45\textwidth} \caption{
(Color online)
Top: Panel A shows the 1 ML pseudomorphic state, panels B and C are 
two 
fcc-hcp domain configurations obtained from the AM procedure. 
The panels A, B and C show
the $15\times 30$ overlayer (light color) and the next-nearest
substrate layer (dark color), as in Fig. \ref{fig:plates}.
Bottom: Minimum energy paths connecting the pseudomorphic state 
to fcc-hcp domain configurations.}
\label{fig:paths}
\end{figure}

{\it Monolayer regime. \--} Although we find that pseudomorphic
Cu at 1 ML coverage, 
which is shown in Fig.\ref{fig:paths}, panel A, is stable 
upon local energy minimization, both GO and AM
methods can locate lower energy structures. They consist of
alternating fcc and hcp stacking domains similar to the submonolayer
islands. We will refer to these structures collectively as the
fcc-hcp domain configurations (cf. Fig. 2).
We are thus led to investigate what is the stability of the
pseudomorphic film relative to the fcc-hcp domains and to what
extent the adsorbate is epitaxial.

To mimic the row dislocation networks sometimes reported from
similar systems \cite{Schaff01}, we perturb the $15\times30$
simulation cell by pushing a $x$ directional row of the Cu layer
$0.05$ \AA $\;$ down to the negative $z$ direction. 
We activate the system with only 3 layers of Pd, but after 
finding a transition path, we increase the substrate 
thickness to 9 by adding 6 layers below the originals. 
As before, the two bottom of the enlarged system 
are kept frozen. The AM then yields a configuration where the 
overlayer has contracted and split to reveal an $x$ directional row 
from the substrate, see Fig. \ref{fig:paths}, panel B. 
This row dislocation state has lower energy than the pseudomorphic 
state and contains fcc-hcp domains, but the energy
barrier exceeds $1$ eV.

The fcc-hcp domains both in the submonolayer islands and from the GO
runs at 1 ML seem to have a fairly irregular structure. To estimate
the energy barrier connecting the pseudomorphic layer to these
structures, we pushed $15$ randomly selected Cu atoms $0.05$ \AA 
$\;$ to the nearest $y$ edge direction. The result after the AM procedure
also shows fcc-hcp domains, see Fig. \ref{fig:paths}, panel C. These
domains have striking similarity to what is described in Refs.
\onlinecite{Carter95,Hamilton95, Gabaly05, Hwang97, Figuera01,
Schmid97, Gunther95}. The energy path connecting the pseudomorphic
state to the fcc-hcp domain is characterized by multiple large
barriers which exceed the row dislocation barrier. The holes on the
overlayer are now better aligned to allow stress release in several
directions (cf. Fig. 2, panels B and C). Because the row dislocation
shows the stacking domains and the energy barrier is in the same
range as in the disordered case, it is possible to view it as a
special case which also belongs to the fcc-hcp domain class.

\begin{figure}[ht]
\begin{center}
\epsfig{file=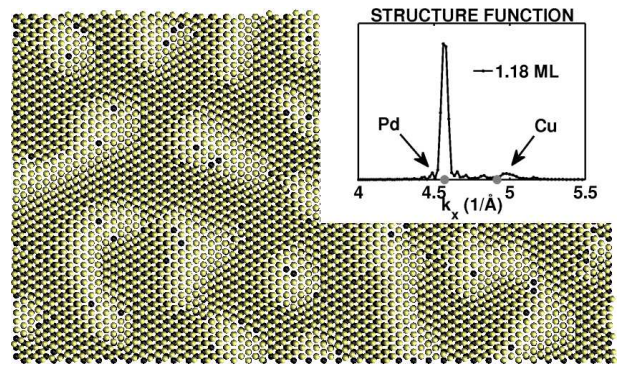, width=0.45\textwidth}
\caption{\label{structfact} 
(Color online)
Randomly placed 0.18 ML adatoms on top of a pseudomorphic Cu layer
incorporate spontaneously and nucleate a fcc-hcp stacking 
domain structure to the interfacial layer. The colors are as in Fig. \ref{fig:plates}. The sparse black spots are holes in the interfacial layer. 
Insert: The average structure function in the $x$ direction
for eight realizations of the adsorbate layer after random
incorporation, all with the same nominal coverage 1.18 ML.
The Pd-like peak, which is ideally above the left grey spot, is 
strong because the interior of both fcc and hcp domains has 
Pd-like lattice constant. The peaks near the ideal Cu-position 
(right grey spot) are related to the domain walls, where most of 
the stress relaxation takes place.
}
\end{center}
\end{figure}

The NEB energy paths of Fig. \ref{fig:paths} suggest that if a
uniform pseudomorphic overlayer can be formed, high energy barriers
will prevent the decay into the fcc-hcp domain structure. If this
overlayer was grown with conventional deposition methods, some
adatoms would eventually land on the film. Therefore we are
led to ask if a few adatoms on a single pseudomorphic layer
would be enough to activate a defect nucleation.

In our initial state we placed $N_{\rm ad}$ Cu adatoms, all at
the same time, on top of a pseudomorphic layer at randomly selected
fcc stacking positions. Substrate layer sizes $15\times 30,$
$30\times 60$ and $40\times 80$ were used with increasing $N_{\rm
ad}$. We find that with simple energy minimization
$N_{\rm inc}$ spontaneously incorporate into the interfacial layer.
For all the cases considered $N_{\rm inc}$ saturates 
near the number difference 
between the fully pseudomorphic and fully
relaxed layers, which is roughly $2f$ times the layer size.
The remaining Cu adatoms stay in the topmost layer. As a result of
the incorporation, a fcc-hcp domain structure is again formed, 
as shown in Fig.\ref{structfact}. In saturation, the fcc stacking is
slightly more frequent than hcp. GO runs also display similar
low-energy patterns. Overlayer structures with fcc-hcp domains
similar to those in Fig. \ref{fig:plates} have been observed on various
heteroepitaxial systems on (111) surfaces \cite{Hamilton95}.

{\it Discussion. \--} The results presented here are in agreement
with the experimentally reported layer-by-layer growth of Cu on Pd(111)
\cite{Oral90,Liu99,deSiervo05A}, since flat islands are the low
energy configurations in the whole submonolayer range considered
here. However, our results predict that the most favorable
arrangement of Cu on top of Pd(111) is the fcc-hcp domain
configuration rather than an epitaxial configuration.
This result was obtained both from the global optimization method as
well as from the activated and spontaneous relaxation starting from
a perfect epitaxial state.

The experimental peak intensities from Au\-ger electron spectroscopy
(AEG), low energy electron spectroscopy (LEED) \cite{Oral90} and
reflection high energy electron dif\-frac\-tion (RHEED)
\cite{deSiervo05A} were interpreted to indicate that the Cu
deposition is epitaxial up to $2-3$ ML. In contrast, in our
calculations fcc-hcp domain configurations occur already at the
submonolayer regime.
In the measured RHEED intensities, however, there is a Cu-like peak
already visible and distinguishable from the Pd-like 
at $\approx 1.5$ ML \cite{deSiervo05A}. 
To estimate the degree of epitaxy in the overlayers 
exhibiting fcc-hcp domains, 
we plot the average structure function $|\sum_{\rm Cu} \exp(ik_x x)|^2$ 
averaged over eight configurations 
after incorporation at 1.18 ML coverage in Fig.\ref{structfact}.
The averaging is done to remove the dependence 
on the random adatom positions. 
Most of the signal is Pd-like because both the fcc and hcp domains 
are nearly coherent with the substrate.
However, there is some relaxation originating from the domain 
walls and this produces two peaks near the ideal Cu-like location, 
corresponding to distorted heavy and superheavy walls. 
After averaging the Cu-peaks are less visible and
for realistic temperatures the fluctuations would further 
smear the data. If the Pd signal slope were changing faster than the
slope of the Cu signal as a function of the wave vector length $k_x,$ 
the latter signal would be difficult to resolve.
However, there is no reason why the morphology could not 
change to completely relaxed layers from the domain structure after
deposition of a few layers, as the experimental data shows.
Thus, in our view, the existence
of the stacking domains is not ruled out by the experiments.

Similar fcc-hcp domains have been observed experimentally
in Cu/Ru(0001) with misfit $-5.5\%$ \cite{Gabaly05, Hwang97,
Figuera01, Schmid97, Gunther95}. Theoretical studies suggest that
the existence of the structure is not sensitive to details of the
stressed fcc(111) system \cite{Carter95, Hamilton95}. Instead, these
structures arise from the existence of two favorable states, fcc and
hcp sites, and from the competition between strain energy relief and the
energy cost for dislocation formation. The loss of epitaxy of
Cu/Ru(0001) occurs above 1 ML coverage but in our system with misfit
$-7\%$ the critical coverage is below $1$ ML, which is in line with
the higher stress level. In the Cu/Pd(111) system the incorporation
would not lead to further stress release because the layer is
already patterned below $1$ ML. However, in the Cu/Ru(0001) system
the process is visible. The XPS intensity 
data for Cu/Pd(111) system
also leave room for layer-by-layer morphology above $1$ ML
\cite{deSiervo05A}.

{\it Conclusion. \--} We have used computational approach to
understand the structure of Cu films on the Pd(111) surface in the
low coverage regime extending from $20$-atom islands to $1$ ML.
Our results agree with layer-by-layer growth, as is observed
experimentally. Our model predicts that the epitaxiality is lost
already below $1$ ML coverage and Cu forms hcp and fcc stacking
domains, separated by wall regions in which Cu recovers its
lattice spacing and releases its tensile stress. This pattern,
similar to what is seen in Cu/Ru(0001) at higher coverages,
encourages further experimental investigations directly probing the
overlayer structure.

{\it Acknowledgements. } \-- This work has been supported in part by
the Academy of Finland through the COMP CoE grant, CSC-IT Center of
Science Ltd. through allocation of computer resources, 
and joint funding under EU STREP Grant No. 016447 MagDot 
and NSF DMR Grant No. 0502737.  J.J. acknowledges 
support from the Finnish Cultural Foundation.
E.G. was supported by Funda\c{c}\~ao de Amparo \'a Pesquisa do Estado de S\~{a}o Paulo - FAPESP (Grant no. 07/08492-9).


\begin{thebibliography}{00}

\bibitem{Gabaly05}
F. El Gabaly, W. L. W. Ling, K. F. McCarthy and J. de la Figuera,
Science {\bf 27}, 1303 (2005).
\bibitem{Hwang97}
R. Q. Hwang and M. C. Bartelt, Chem. Rev. {\bf 97}, 1063 (1997).
\bibitem{Figuera01}
J. de la Figuera, A. K. Schmid, N. C. Bartelt, K. Pohl and R. Q.
Hwang Phys. Rev. B {\bf 63}, 165431 (2001).
\bibitem{Schmid97}
A. K. Schmid, N. C. Bartelt, J. C. Hamilton, C. B. Carter and R. Q.
Hwang Phys. Rev. Lett. {\bf 78}, 3507 (1997).
\bibitem{Gunther95}
C. G\"{u}nther, J. Vrijmoeth, R. Q. Hwang and R. J. Behm, Phys.
Rev. Lett. {\bf 74}, 754 (1995).
\bibitem{Carter95}
C. B. Carter and R. Q. Hwang, Phys. Rev. B {\bf 51}, 4730 (1995).
\bibitem{Hamilton95}
J. C. Hamilton and S. M. Foiles, Phys. Rev. Lett. {\bf 75}, 882
(1995).
\bibitem{Pushpa03} R. Pushpa and S. Narasimhan,
Phys. Rev. B {\bf 67}, 205418 (2003).
\bibitem{Ubiq}
In fcc(111) geometry the simplest defect arrangements whose
Burgers vectors cancel at distance are pairs 
pointing to opposite directions or triplets
radiating from a common centre. 
The meeting points of partial dislocations belong to
either of these categories and the general features of the resulting domain 
structure can be expected to be insensitive to details of the interactions. 
\bibitem{alloying}
Annealing the Cu/Pd(111) system above $450$ K leads to formation of
a surface alloy \cite{Liu99,deSiervo05B}, which is more stable than
the pseudomorphic film. At room temperature this is not
seen and is supposedly behind a considerable kinetic barrier, in
which case the local stress release processes considered here have a
lower activation energy.
\bibitem{Liu99} G. Liu, T. P. St. Clair and D. W. Goodman,
J. Phys. Chem. B \textbf{103}, 8578 (1999).
\bibitem{deSiervo05A} A. de Siervo, R. Paniago, E.A. Soares, H.-D. Pfannes, R. Landers and G.G. Kleiman,
Surf. Sci. \textbf{575}, 217 (2005).
\bibitem{Oral90} B. Oral and R.W. Vook,
J. Vac. Sci. Technol. A {\bf 8}, 3048 (1990).
\bibitem{deSiervo05B} A. de Siervo, E. A. Soares, R. Landers and G. G. Kleiman,
Phys. Rev. B {\bf 71,} 115417 (2005).

\bibitem{Daw84} M. S. Daw and M. I. Baskes,
Phys. Rev. B {\bf 29}, 6443 (1984).
\bibitem{Foiles86}  S. M. Foiles, M. I. Baskes and M. S. Daw,
Phys. Rev. B {\bf 33}, 7983 (1986).
\bibitem{Lu05} Y. Lu, M. Przybylski, O. Trushin, W. H. Wang, J. Barthel, E. Granato, S. C. Ying and T. Ala-Nissila
Phys. Rev. Lett. {\bf 94}, 146105 (2005).

\bibitem{Chen03} Z-X. Chen, K. M. Neyman, A. B. Gordienko, and N. Rosch,
Phys. Rev. B {\bf 68}, 075417 (2003).
\bibitem{Mottet02} C. Mottet, G. Tr\'{e}glia and B. Legrand,
Phys. Rev. B {\bf 66}, 045413 (2002).
\bibitem{Rossi09} G. Rossi and R. Ferrando, 
J. Phys.: Condens. Matter. {\bf 21,} 084208 (2009).
\bibitem{Wales03book} D. J. Wales,
in {\it Energy Landscapes with Applications to
Clusters, Biomolecules and Glasses}, (Cambridge University Press,
Cambridge, 2003).
\bibitem{Jalkanen05} J. Jalkanen, O. Trushin, E. Granato, S.C. Ying and T. Ala-Nissila,
Phys. Rev. B {\bf 72,} 081403(R) (2005).
\bibitem{Trushin09} O. Trushin, J. Jalkanen, E. Granato, S. C. Ying and T. Ala-Nissila,
J. Phys.: Condens. Matter {\bf 21},  084211 (2009).
\bibitem{tru02b} O. Trushin, E. Granato, S.-C. Ying, P. Salo and T. Ala-Nissila,
Phys. Stat. Sol. B, {\bf 232},100 (2002).
\bibitem{tru04} O.S. Trushin, P. Salo, T. Ala-Nissila, and S.C. Ying,
Phys. Rev. B. {\bf 69}, 033405 (2004).
\bibitem{neb} H. J\'onsson, G. Mills and K. W. Jacobsen,
in {\it Classical and Quantum Dynamics in Condensed Phase
Simulations}, ed. by B. J. Berne {\it et al} (World Scientific,
Singapore, 1998).

\bibitem{Schaff01} O. Schaff, A. K. Schmid, N. C. Bartelt, J. de la Figuera and R. Q. Hwang,
Mater. Sci. Eng., A {\bf 319},  914 (2001)

\bibitem{Nielsen83} O. H. Nielsen and R. M. Martin, 
Phys. Rev. Lett. {\bf 50,} 697 (1983).
\bibitem{Vitek87} V. Vitek and T. Egami,
Phys. Stat. Sol. (b) \textbf{144}, 145 (1987).
\bibitem{Ouahab05} A. Ouahab, C. Mottet and J. Goniakowski,
Phys. Rev. B \textbf{72}, 035421 (2005).

\bibitem{Zeppenfeld88} P. Zeppenfeld, K. Kern, R. David and G. Comsa,
Phys. Rev. B {\bf 38,} 3918 (1988).
\end{thebibliography}
\end{document}